\newif\ifarxiv
\arxivtrue

\ifarxiv
\documentclass{article}
\usepackage{arxiv}
\usepackage[square,sort,comma,numbers]{natbib}
\usepackage{graphicx}
\usepackage{booktabs}
\usepackage{hyperref}
\usepackage{authblk}
\else
\documentclass[manuscript,screen]{acmart}
\AtBeginDocument{%
  \providecommand\BibTeX{{%
    \normalfont B\kern-0.5em{\scshape i\kern-0.25em b}\kern-0.8em\TeX}}}
\setcopyright{none}
\acmConference[Conference acronym 'XX]{Make sure to enter the correct
  conference title from your rights confirmation emai}{June 03--05,
  2018}{Woodstock, NY}
\acmPrice{15.00}
\acmISBN{978-1-4503-XXXX-X/18/06}
\fi

\usepackage[inline]{enumitem}
\usepackage{graphicx}
\usepackage{subfig}
\usepackage{tabularx}
\usepackage{enumitem}
\usepackage{soul}
\soulregister{\section}{1}
\soulregister{\subsection}{1}

\ifarxiv

\title{Through the Looking-Glass: Transparency Implications and Challenges in Enterprise AI Knowledge Systems}

\author[1]{Karina Cortiñas-Lorenzo}
\author[1]{Siân Lindley}
\author[2]{Ida Larsen-Ledet}
\author[1]{Bhaskar Mitra}

\affil[1]{Microsoft \\\texttt{\small \{karinac, sianl, bmitra\}@microsoft.com}}
\affil[2]{University College Cork \\\texttt{\small ilarsen-ledet@ucc.ie}}

\else

\title[Through the Looking-Glass: Transparency in Enterprise AI Knowledge Systems]{Through the Looking-Glass: Transparency Implications and Challenges in Enterprise AI Knowledge Systems}

\author{Karina Cortiñas-Lorenzo}
\affiliation{%
  \institution{Microsoft}
  \city{Zurich}
  \country{Switzerland}
  }
\email{karinac@microsoft.com}
\author{Siân Lindley}
\affiliation{%
  \institution{Microsoft Research}
  \city{Cambridge}
  \country{United Kingdom}}
\email{sianl@microsoft.com}
\author{Ida Larsen-Ledet}
\affiliation{
  \institution{School of Applied Psychology, University College Cork}
  \city{Cork}
  \country{Ireland}}
\email{ilarsen-ledet@ucc.ie}
\author{Bhaskar Mitra}
\affiliation{%
  \institution{Microsoft Research}
  \city{Montreal}
  \country{Canada}}
\email{bmitra@microsoft.com}

\fi

\begin{document}

\ifarxiv
\maketitle
\fi

\begin{abstract}
   Knowledge can't be disentangled from people. As AI knowledge systems mine vast volumes of work-related data, the knowledge that's being extracted and surfaced is intrinsically linked to the people who create and use it. When predictive algorithms that learn from data are used to link knowledge and people, inaccuracies in knowledge extraction and surfacing can lead to disproportionate harms, influencing how individuals see each other and how they see themselves at work. In this paper, we present a reflective analysis of transparency requirements and impacts in this type of systems. We conduct a multidisciplinary literature review to understand the impacts of transparency in workplace settings, introducing the looking-glass metaphor to conceptualize AI knowledge systems as systems that reflect and distort, expanding our view on transparency requirements, implications and challenges. We formulate transparency as a key mediator in shaping different \emph{ways of seeing}, including \emph{seeing into} the system, which unveils its capabilities, limitations and behavior, and \emph{seeing through} the system, which shapes workers' perceptions of their own contributions and others within the organization. Recognizing the sociotechnical nature of these systems, we identify three transparency dimensions necessary to realize the value of AI knowledge systems, namely system transparency, procedural transparency and transparency of outcomes. We discuss key challenges hindering the implementation of these forms of transparency, bringing to light the wider sociotechnical gap and highlighting directions for future Computer-supported Cooperative Work (CSCW) research.
\end{abstract}

\section{Introduction}

\begin{quote}
\emph{Oh, Kitty! how nice it would be if we could only get through into Looking-glass House! I’m sure it’s got, oh! such beautiful things in it! Let’s pretend there’s a way of getting through into it, somehow, Kitty. Let’s pretend the glass has got all soft like gauze, so that we can get through.}

\emph{...Oh, what fun it’ll be, when they see me through the glass in here, and can’t get at me!}

\begin{flushright} 
-- Lewis Carroll, “Through the Looking-Glass”
\end{flushright}
 
\end{quote}

Knowledge is created, used and shared by people. Inherently human-centric and relational, in organizations it becomes embedded not only in documents and repositories, but also in processes, norms, communications and actions \cite{davenport1998working, orlikowski2002knowing, lindley2023building, thomas2001knowledge, nonaka1994dynamic}. To foster effective utilization of knowledge in an organization, \emph{Artificial Intelligence (AI) knowledge systems} operate by mining vast volumes of work-related data, automatically extracting and surfacing knowledge to help individuals progress work \cite{winn2021enterprise, microsoftviva}. Because the production and the use of knowledge is embedded in social phenomena \cite{thomas2001knowledge}, these systems are intrinsically sociotechnical and must be human-centered by design. 

One way to extract knowledge from work-related data involves mining associations between knowledge areas, artefacts and individuals \cite{wilkins2020designing, vivatopics}. For example, this could involve automatically identifying who knows about a given knowledge topic in a company or linking related documents to a given subject area. As AI knowledge systems extract and surface these associations, they enable and shape different types of seeing \cite{larsen2022ways, wolf2016seeing}, impacting how we see and perceive others in an organization and how we see ourselves and our own contributions to organizational knowledge. In the enterprise context, what information gets brought to the foreground by these systems and what information is pushed to the periphery can influence who we interact and collaborate with, what information is trusted and how social interactions are established, potentially leading to interpersonal harms if the system misses relevant associations, surfaces inaccurate information or exposes work-related data without explicit user consent and proper contextualization \cite{shelby2023sociotechnical, larsen2022ethical, gausen2023framework}. As others are seen, individuals might also wonder how they are seen by others \emph{through} the system lenses, what aspects of their work get surfaced by the system, to whom and where. How these questions are addressed and what information is disclosed when doing so has the potential to impact self-concept beliefs, influencing the extent to which these systems can augment and amplify humans in creating, sharing and applying knowledge in practice.

As we design and develop systems aimed at improving knowledge workers' efficiency and productivity, what gets seen and how end-users interpret what they see can have deep implications in the workplace, potentially leading to worsened situations by hindering individual's efficacy and well-being. In this context, transparency can help end-users of AI knowledge systems understand, interpret and calibrate the importance of what they see \emph{through} the system, helping mitigate potential negative effects on self-concept beliefs and perceptions of others \cite{french2018algorithmic, lee2022algorithmic, wolf2020knowledge}. Yet, implementing transparency in complex real-world scenarios remains a challenge \cite{eiband2018bringing, liao2023ai}. As novel AI approaches such as Large Language Models (LLMs) continue to be developed, several challenges arise, such as opaque and large architectures, complex and uncertain AI capabilities and new stakeholder groups with unique transparency needs \cite{liao2023ai}. In enterprise knowledge systems, emerging LLM-based approaches can also exacerbate and pose new risks \cite{gausen2023framework}, introducing new biases into existing information retrieval capabilities \cite{dai2024bias, bommasani2021opportunities} and magnifying existing risks such as exposure inequality \cite{fabbri2022exposure, amendola2024understanding, singh2018fairness, anthis2024impossibility}. As AI capabilities continue to evolve, several voices have raised concerns about the widespread deployment of LLM-infused applications before appropriate safeguards and risk mitigations are in place, highlighting the importance of taking additional measures to ensure that transparency and other responsible AI considerations remain a priority \cite{liao2023ai, kenthapadi2023generative, humphreys2024ai}. In this context, while past research has recognized transparency as a crucial aspect of responsible design \cite{larsen2022ethical, lindley2023building, gausen2023framework}, it's unclear how transparency should be implemented in enterprise AI knowledge systems, which forms of transparency should be prioritized and what the potential impacts and challenges related to their implementation might be. As AI systems continue to grow in complexity, reflecting on how to implement AI transparency in enterprise knowledge systems and the ethical implications of doing so is crucial to ensure these systems are human-centered by design and can augment both human labor and well-being.

To this end, this paper presents a reflective analysis of the potential requirements and impacts of AI transparency in enterprise AI knowledge systems. By \emph{AI knowledge system}, we understand any system that extracts and surfaces knowledge by using predictive algorithms that learn from data. To inform our analysis, we conducted a multidisciplinary literature review drawing from different fields including computer science, psychology, organizational studies and human-computer interaction. The findings highlighted the complexities of incorporating transparency in enterprise settings, helping us contextualize the impacts of AI transparency in knowledge systems. We then hypothesized that transparency in these systems can shape different \emph{ways of seeing} in the workplace. Drawing on Lewis Carroll's depiction of a magical mirror \cite{carrol2006through}, we introduced the \emph{looking-glass} metaphor as a conceptual tool for practitioners to critically examine the needs and implications of different transparency dimensions. As compared to other metaphors used to describe AI systems, the looking-glass metaphor conveys the idea that AI systems not only reflect, but also distort and reshape what is perceived, often in ways that are imperceptible to users, leading them to accept skewed perceptions as truth. The proposed metaphor is particularly valuable in sensitive contexts like workplace settings, where transparency can shape perceptions by creating a false sense of agency \cite{ananny2018seeing}, impacting power dynamics in an organization \cite{gausen2023framework} or introducing tensions between different stakeholder groups \cite{park2022designing}. By broadening the scope of transparency, the looking-glass metaphor facilitates wider discussions of its potential effects, helping practitioners overcome ``failures of imagination'' \cite{boyarskaya2020overcoming} when anticipating transparency impacts and challenges. We demonstrated the use of the metaphor by applying it to our analysis of transparency implications in AI knowledge systems. This analysis highlighted the limitations of transparency initiatives solely focused on explaining the inner workings of AI knowledge systems (system transparency), emphasising the need to also consider how the system operates in the organizational context (procedural transparency) and the outcomes of system use in the long run (transparency of outcomes). As we reflected on the wider sociotechnical gap or divide between what we must support socially and what we can support technically \cite{ackerman2000intellectual}, we used the term \emph{AI transparency} to refer to a combination of decision-making and tooling aimed at revealing and concealing information about an AI system.

In summary, our contributions are four-fold: 

\begin{itemize}
\item Building on past literature, we present the looking-glass metaphor and use it to conceptualize AI knowledge systems as systems that reflect and distort, enabling different types of seeing. This framework allows us to widen the scope of transparency needs, moving from a narrow system focus where the sole goal is allowing \emph{seeing into} the system, to a wider view where what lies outside the system is brought to light.
\item We use the looking-glass metaphor to explore the impacts of different types of seeing and identify key requirements for transparency in AI knowledge systems. Recognizing the sociotechnical nature of these systems, we identify three transparency dimensions that we believe are necessary to realize the value of AI knowledge systems, namely system transparency, procedural transparency and transparency of outcomes.
\item We discuss key challenges that hinder the implementation of these forms of transparency, bringing to light the wider sociotechnical gap and tradeoffs to enable decision-making and support interdisciplinary discourse.
\item We identify future areas of transparency research in AI knowledge systems, encouraging Computer-supported Cooperative Work (CSCW) researchers to further study transparency implications and first-order approximations to the problems we're highlighting.
\end{itemize}

The analysis we contribute focuses on AI transparency requirements, impacts and challenges specifically in relation to enterprise knowledge systems. It is aimed at informing further empirical research and encouraging the CSCW community to reflect on the broader social, ethical and technical impacts of AI transparency in workplace settings.

\section{Related Work}

\subsection{Algorithmic Impact Assessments}

The analysis of AI impacts is considered a critical step to understand the potential implications of AI systems in the world \cite{boyarskaya2020overcoming, watkins2021governing, reisman2018algorithmic, metcalf2021algorithmic}. Extensive research has shown that AI systems can lead to unintended consequences, having negative effects at the individual, collective and societal levels in ways that are often difficult to predict or control \cite{barocas2023fairness, amodei2016concrete, selbst2018intuitive, binns2018fairness}. In response to these concerns, \emph{algorithmic impact assessments} aim to anticipate adverse impacts \emph{before} they occur (i.e., \emph{``ex ante''} \cite{watkins2021governing}), prompting interventions in the system's design to mitigate any potential negative effects. Drawing directly from environmental protection, privacy and human rights policy domains \cite{reisman2018algorithmic}, impact assessments are typically measured against alternative scenarios, making use of open-ended questions to uncover unknown unknowns \cite{selbst2021institutional} instead of ``checklists'' \cite{boyarskaya2020overcoming}, and often relying on prior use cases of technology and narrative records of how the system was designed and iteratively developed \cite{watkins2021governing}. By drawing attention to design choices and consequences \cite{watkins2021governing}, impact assessments can influence decision-making, encouraging more deliberate and thoughtful designs \cite{watkins2021governing}. However, determining what constitutes an impact is not always clear \cite{metcalf2021algorithmic}. Because impacts are not directly observed nor measured, the assessment of impacts is considered as much of a ``thought experiment as an empirical endeavor'' \cite{metcalf2021algorithmic}, being prone to ``failures of imagination'' \cite{boyarskaya2020overcoming}, where possible impacts may be ignored or underestimated.

While impact assessments may not fully capture all possible implications, the discussions they generate are critical spaces for expressing dissent and facilitating negotiation \cite{watkins2021governing}. Because foreseeing failures that have not been observed before is inherently a difficult task, wider and interdisciplinary discussions are needed \cite{boyarskaya2020overcoming}, also for AI work aimed at making AI systems fair and transparent \cite{olteanu2023responsible}. While transparency in AI systems is often considered a key mitigation strategy on its own and a fundamental dimension for responsible AI design \cite{jobin2019global, liao2023ai, diaz2023connecting}, in contexts such as the enterprise it can have multiple ramifications and adverse consequences \cite{gausen2023framework, park2022designing}, requiring a more in depth analysis of its potential impacts. In this paper, we fill this gap by critically reflecting on the needs for transparency in AI knowledge systems and the possible risks and challenges related to its implementation in the workplace.

\subsection{AI Metaphors}

Ideas about what technology is, or should do, can influence how its impacts are assessed and the effects it has on the world \cite{gilbert2023reframing}. As cognitive tools \cite{lakoff2008metaphors}, metaphors play a crucial role in shaping these ideas, influencing both design, development and evaluation choices. Considered a generative design tool \cite{logler2018metaphor} and pervasive in everyday life \cite{lakoff2008metaphors}, metaphors can provide new ways of seeing, thinking about and experiencing technology, illuminating and hiding different aspects of it and creating opportunities to view objects or phenomena as something different \cite{murray2022metaphors}. Past research has shown that the practice of AI draws heavily on metaphors \cite{murray2022metaphors}. Some unnoticed foundational metaphors related to AI transparency include ``explanation'', relating the computational information about a model operation to the social processes by which humans construct explanations, and ``closed-box''\footnote{We use the term ``closed-box'' instead of ``black-box'' to avoid perpetuating the narrative that ``black'' is undesirable, wrong or something to be avoided}, implying that opening the opaque box would expose what's hidden and divert attention from asking who lies outside the box or how the box came to be \cite{murray2022metaphors}. Metaphors like these can not only shape how AI systems are designed but also influence the expectations and evaluations of end-users when interacting with these systems \cite{khadpe2020conceptual}. For example, past work has conceptualized AI systems as ``collaborators'', ``teammates'', ``assistants'' or ``co-pilots'' \cite{khadpe2020conceptual, sarkar2023enough, sellen2024rise, sarkar2024copilot}. While some of these metaphors can personify AI systems with problematic consequences \cite{khadpe2020conceptual, sarkar2023enough}, others can illuminate new design spaces, shaping the ways we interact with new technology \cite{sellen2024rise}.

Different metaphors have been used to describe AI personalization systems that implicitly or explicitly represent people. For instance, \citet{french2018algorithmic} and \citet{hess2014you} conceptualize AI recommenders as mirrors reflecting the system's approximation of the self back to the individual in the form of personalized content. Conceiving algorithmic personalization as a form of feedback about the self \cite{french2018algorithmic}, interacting with AI recommenders is seen as ``looking into a mirror rather than looking out a window'' \cite{hess2014you}. Personalized AI algorithms in digital platforms such as TikTok have also been conceptualized as ``crystals'' \cite{lee2022algorithmic}, conveying the idea that a crystal can be polished or refined by the user and inspiring new design choices that emphasize user agency and autonomy. The image of a crystal helps understand the capabilities of the system to support reflection of end-user self-concepts via personalized content, as well as refraction, shaping perspectives of others through the algorithm \cite{lee2022algorithmic}. In this paper, we build on this work and propose the looking-glass metaphor to conceptualize AI knowledge systems. As our goal is to reflect on the needs for transparency in these systems and the challenges and impacts arising from its implementation, the metaphor highlights how AI systems can distort reality in subtle ways, skewing perceptions of the self and others and mediating transformative effects.

\section{Method}

\subsection{Literature Review}

To guide our analysis, we reviewed multidisciplinary literature, focusing primarily on computer science, psychology, organizational studies and human-computer interaction. Due to the multidisciplinarity of the reviewed works, we reviewed the literature iteratively, continuously refining our search and selection criteria as new insights emerged. To ensure multidisciplinary coverage, papers were drawn from multiple sources, including Google Scholar, ACM Guide to Computing Literature, PsycINFO and SAGE Journals. Weekly discussions among the authors helped synthesize the insights and refine the search, collaboratively iterating the emerging themes.

The literature review focused on three key aspects necessary for the assessment of transparency requirements and impacts in enterprise AI knowledge systems: the conceptualization of transparency, the use of AI methods in these systems and the impacts of AI outputs on self-perception and social processes. These aspects guided the search and the selection of relevant references across multiple disciplines. First, we reviewed the conceptualization of transparency, unpacking what is meant by the term in different contexts, both technical and social. By exploring different conceptualizations, we also sought to identify where tensions and trade-offs might arise. Second, we examined the use of AI methods in enterprise knowledge systems taking a narrative approach, reviewing their introduction over time and the different CSCW perspectives shaping their role in knowledge extraction and surfacing. Lastly, we explored the reported impacts of AI systems on self-perception beliefs and social processes to understand the extent to which AI transparency, and the lack thereof, can impact individuals in workplace settings. By reviewing these three interconnected areas, the literature review aimed to inform our understanding of the nuances of AI transparency in enterprise contexts, helping inform our assessment of requirements and potential impacts.

\subsection{Reflective Analysis and Positionality Statement}
As researchers, we acknowledge that our backgrounds, life experiences and beliefs can shape our work. Indeed, the reflective analysis we contribute in this paper was motivated by our past work researching how AI technology can augment knowledge workers and be integrated in enterprise settings in a way that's ethical and human-centered. Our analysis of potential transparency requirements and challenges was influenced by our own lived experiences as end-users of AI knowledge systems within a large technology company. As we reflected on the types of seeing enabled by these systems and how these can be shaped by transparency mechanisms, we combined reflections from the literature review with insights gathered via discussions with AI researchers working in AI knowledge systems in the same company, who provided valuable inputs on the practical challenges of implementing different forms of transparency. Acknowledging the role that the researcher's position plays in the creation of knowledge, our research team engaged with reflexive practice. Our educational background and collective experience in HCI and machine learning, along with our roles as knowledge workers in a large technology corporation, influenced our own understanding of what it means to be a knowledge worker, how an organisation works and what constitutes an adverse impact. We recognize, however, that these perspectives are not unique and are also shaped by our location in a Western context, influencing our interpretations. As normative decisions were made throughout the analysis, we acknowledge that the foresight of transparency requirements and impacts presented in this paper is necessarily incomplete and not value neutral.

\section{Literature Review}

We begin by reviewing the use of the term \emph{transparency} in different disciplines and the tensions arising from different conceptualizations. Next, we characterize AI knowledge systems in work settings and review the use of AI algorithms to extract and surface organizational knowledge. We finish by exploring the self-concept psychology and sociology theories explaining the mechanisms that underlie self-perception and how the formation of beliefs about oneself can be shaped by AI systems.

\subsection{Transparency: \emph{seeing} and \emph{knowing}}
As AI systems continue to evolve and become integrated into several aspects of our lives, calls for greater transparency have been widespread. However, what is meant by transparency is often left implicit \cite{andrada2023varieties, corbett2023interrogating}. While in fields such as optics and physics transparency denotes the property of a material to let light pass through, in others it's often conceptualized as a wide array of 
practices aimed at holding collectives accountable \cite{larsson2020transparency}. Regardless of the discipline, most works see transparency as being closely intertwined with \emph{seeing}: a transparent object enables observation through it whilst a transparent organization makes aspects of its operations visible and, hence, open to scrutiny. Seeing, understood in a metaphorical sense as ``being aware of'', doesn’t automatically equate to the kind of knowing or understanding that enables action \cite{ananny2018seeing, larsson2020transparency}. If we understand transparency as informational transparency, that is, as information about an AI system that should be disclosed to enable appropriate understanding \cite{liao2023ai}, then, considering the sheer and increasing complexity of AI systems, transparency necessarily requires a conscious decision and assessment of which information \emph{should} be made visible, sometimes at the expense of concealing other information \cite{corbett2023interrogating}. 

What to conceal, what to reveal and how much to conceal or reveal are timeless questions that have been examined in philosophy, computing sciences and Human Computer Interaction (HCI) communities. In philosophy, past works have differentiated between \emph{transparency in use}, understood as technology being so easy to use that it allows individuals to \emph{see through} it and focus on the activity at hand \cite{bodker2021through}, and \emph{reflective transparency}, interpreted as the extent by which a system allows a user to \emph{see into} its mechanisms \cite{andrada2023varieties, perez2022situating, clowes2020internet}. In CSCW, the discussion was initially framed from an awareness perspective \cite{mantau2022awareness}, with a focus on the interpretation of computing-based representations of people, activities or artifacts. In representing people and their work, \emph{we have to make a choice, as to what we exclude and include, what we represent in detail and what is abstract, what we actively interpret and what we leave to human interpretation} \cite{chalmers2001information}. In the context of knowledge management, social translucence was defined as an approach to designing digital systems that emphasize sharing social information, supporting visibility, awareness and accountability  \cite{erickson2000social}. Subsequent CSCW work proposed a framework for thinking about social transparency, highlighting potential tradeoffs and implications of making social information visible for collaborative work \cite{stuart2012social}. In HCI, the tension between revealing and concealing information was formalized as a decision on which seams or mismatches to make visible at the expense of what to hold invisible \cite{chalmers2004seamful}, initiating a debate between seamless design -- emphasizing clarity, simplicity and ease of use -- and seamful design, prioritizing configurability, user appropriation and revelation of complexity \cite{inman2019beautiful}. In the realms of privacy research, the question has often been depicted as a conflict between selective control of access to the self, and the desire to see and be seen by others \cite{palen2003unpacking, friedman2007human}. In determining disclosure behavior, privacy works have highlighted the importance of knowing the audiences involved in the process of information sharing and consumption \cite{palen2003unpacking}, as well as the nature of the relationship \cite{joinson2010privacy} and levels of trust \cite{olson2005study} between the discloser and the recipient.

Like privacy \cite{palen2003unpacking}, transparency is not one-sided. More transparency is not necessarily better \cite{corbett2023interrogating, narayanan2023welfarist}. The optimal state along the spectrum of revelation and concealment is highly contextual. By shifting the focus towards the potential impact of making something visible, our work draws on welfarism (the maximization of overall benefits and welfare with consideration to different stakeholders) as a basis to evaluate conflicting transparency desiderata \cite{narayanan2023welfarist}. Building on past research, we understand transparency in the context of AI as a combination of decision-making and tooling aimed at revealing and concealing information, thus modulating different types of seeing in AI systems.

\subsection{Enterprise AI Knowledge Systems}
The vast scale of knowledge generated by organizations, coupled with the challenges of externalizing tacit knowledge and identifying those who hold it, poses several challenges to the efficient utilization of knowledge in work settings. As people work, large volumes of dispersed and siloed data might be generated, such as documents created using productivity tools, meeting transcripts or emails produced via collaboration platforms. In this context, the goal of AI knowledge systems is to support knowledge management and dissemination in an organization by automatically extracting, organizing and surfacing organizational knowledge. In this section, we review the history and role of AI in this type of system, focusing on how AI algorithms can represent individuals both implicitly and explicitly. In what follows, we refer to \emph{AI knowledge systems} as systems that use predictive algorithms that learn from data to support knowledge extraction and sharing.

\subsubsection{Using AI to extract knowledge associations}
\begin{figure}
  \centering
  \includegraphics[scale=0.38]{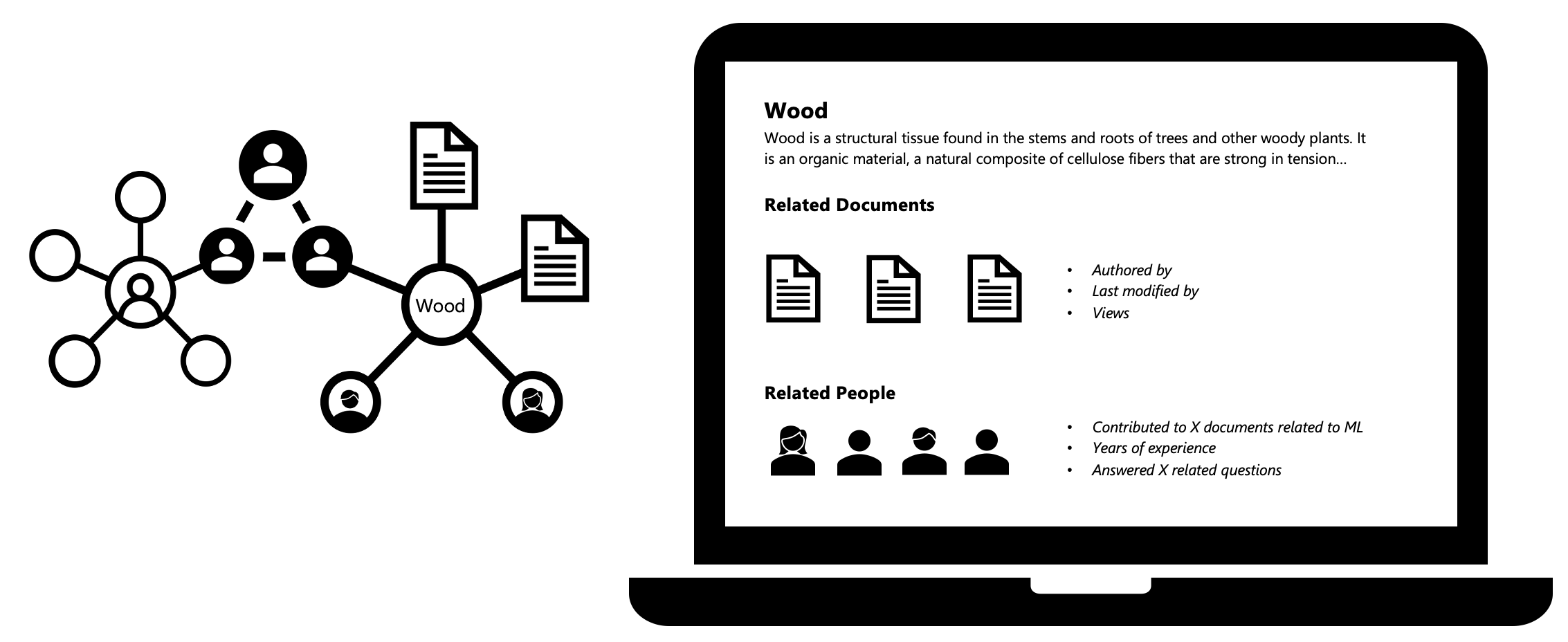}
  \caption{AI knowledge management systems can \emph{implicitly} associate people with topics or areas of knowledge via retrieved documents (e.g., by surfacing a list of related documents) or \emph{explicitly} link people to areas of knowledge (e.g., by displaying a list of related people). When displaying linked documents, various metadata, such as the author, the last individual who modified the document, or the number of views, can be shown. When showcasing a list of related people, metadata such as the number of contributed documents, years of experience in the organization, or the number of questions answered can be displayed.}
  \label{fig:associations}
\end{figure}

Early technical CSCW solutions for knowledge management and dissemination focused on the externalization of knowledge in the form of repositories, emphasizing \emph{knowledge sharing} via the codification of knowledge in databases \cite{ackerman2013sharing}. While these solutions aimed to centralize knowledge, they relied on rigid information structures, requiring users to manually input data and quickly becoming outdated \cite{yimam2003expert, ackerman2000intellectual}. In this context, AI methods have offered a way to address these shortcomings by facilitating the automatic extraction of associations between knowledge areas and knowledge contributions \cite{liebowitz2001knowledge, wilkins2020designing, jarrahi2023artificial}. As such, over the past years, Natural Language Processing (NLP) techniques like those using deep learning and probabilistic models have been applied in different tasks, including information extraction \cite{li2020survey, pawar2017relation, hogenboom2016survey}, automatic knowledge graph construction \cite{zhong2023comprehensive} and knowledge base population \cite{ji2011knowledge, chai2021automatic, winn2021enterprise}. Yet despite continuous AI development and improvement of extraction performance via novel AI approaches such as LLM-based extraction \cite{zhang2024extract, li2024knowcoder, deng2022information}, several voices in the CSCW community have warned against the decontextualization of knowledge, raising awareness of the need to support not only the externalization and categorization of knowledge, but also its social dimensions \cite{lindley2023building, ackerman2000intellectual, davenport1998working}. In this context, \textbf{implicit associations} between knowledge areas and knowledgeable people can help users interpret and contextualize knowledge contributions extracted via AI. For instance, displaying social metadata like the author of a document can allow users to gather more information about the knowledge source, facilitating its reuse in different contexts (Figure \ref{fig:associations}).

Recognizing that knowledge is situated in relation to people, \emph{expertise sharing} efforts in CSCW emphasized the social aspect of knowledge use, placing less emphasis on the externalization of knowledge in favor of fostering collaboration between knowledgeable individuals \cite{ackerman2013sharing, lindley2023building}. A key area of focus became how to identify who knows what in an organization \cite{yimam2003expert, becerra2000role}, finding ways to automatically connect individuals who need information with experts who can provide it \cite{lindley2023building}. While an initial approach involved the use of manually curated expert databases \cite{yimam2003expert}, these solutions suffered from similar shortcomings as knowledge databases, requiring labor intensive maintenance and leading to rapid obsolescence \cite{yimam2003expert, becerra2000role}. As a response to these challenges, different AI methods have been applied in past literature on tasks such as automatic expert profiling in organizations (also referred to as expert profiling or expert identification) \cite{balog2007determining, fazel2011constructing, fazel2011constructing}, people search by expertise on a topic (known as expertise search, expertise retrieval or expert finding systems) \cite{husain2019expert, lin2017survey, gonccalves2019automated} or algorithmic recommendations of knowledgeable people to help progress work \cite{beham2010recommending}. Aimed at fostering collaboration, AI algorithms in these contexts use work-related data to mine \textbf{explicit associations} between areas of expertise and individuals (Figure \ref{fig:associations}). Despite being a construct highly influenced by the social context surrounding an individual \cite{mauksch2020expert}, past literature in the AI domain commonly refers to “expertise” as a modelling target. As such, different sources of data have been used to quantify it, such as documents produced by individuals as they work, completed learning activities, social signal such as answered questions in a forum and endorsements, and/or communications in the form of emails or chats \cite{gonccalves2019automated, lin2017survey, husain2019expert}. Mined associations between artefacts and knowledge areas can also be leveraged when modelling explicit linkages between individuals and knowledge topics. For example, heuristic associations between people and retrieved documents such as authorship or amount of edits can be used to determine who should be linked to a knowledge area. However, when linking individuals to documents, not all people appearing in a document might contribute equally to the topic \cite{lin2017survey}. Thus, how associations between people and content are established and the types of sources considered by the system are important transparency aspects that can help end-users understand system behavior.

\subsubsection{Surfacing knowledge associations}
Once extracted, associations between knowledge and people can be surfaced in a myriad of ways. For example, they can be integrated into search results \cite{vivatopics, glean, bloomfire}, employee profiles \cite{vivatopicsconnections} or served as suggestions in chat-based experiences to help progress work \cite{microsoftcopilotwork, copilotprompt, copilot365, amazonq, atlassian}. As several documents and people might be linked to a given topic of interest, ranking models are commonly used to surface the most associations to end-users \cite{karatzoglou2013learning}. For instance, Viva Topics \cite{vivatopics} presents a curated list of suggested people related to a knowledge topic, allowing end-users to expand the list via UI controls. To identify and rank the most knowledgeable individuals on a given topic, different AI ranking models have been used in past literature, including generative probabilistic models, supervised learn-to-rank methods, similarity approaches and voting methods \cite{lin2017survey, husain2019expert}.

While not all commercialized knowledge systems implement all of these extraction and surfacing capabilities, different knowledge management products offer various combinations of them. For instance, in Microsoft Viva \cite{microsoftviva}, AI technology is used to automatically extract and bring knowledge to users of Microsoft 365 apps via the Topics application, where areas of knowledge, people and knowledge resources are automatically linked and collected into a coherent repository \cite{vivatopics}. Glean \cite{glean} builds a knowledge graph that relates people and content and surfaces this information via search in a personalized way. People are linked to knowledge areas based on whether they've written about a topic. If someone is in need of a knowledge resource that doesn't yet exist, Glean leverages these associations to suggest subject matter experts who could help progress work. Other AI knowledge systems that extract and surface knowledge include different capabilities for different roles in an organization. For example, Bloomfire \cite{bloomfire} provides leadership solutions to help executives and managers drive strategic decision-making, and reporting tools to help knowledge management teams know what resources are being accessed and which ones aren't.

\subsubsection{Risks and ethical considerations} 
\label{subsection:risks}
Extracting and surfacing algorithmic representations of people and their work, whether implicitly or explicitly, raises multiple ethical concerns \cite{larsen2022ethical}, as they can potentially lead to representational, allocative, quality of service and/or interpersonal harms \cite{shelby2023sociotechnical}. Past research has shown that AI information extraction methods often underperform when applied on data that deviates from the training set \cite{abdullah2023systematic, claro2019multilingual}, such as languages with different morphology \cite{humayoun2022urdu, shaalan2019challenges} or low resource levels \cite{hedderich2020survey, deng2022information}, multilingual text \cite{ccetinouglu2016challenges, winata2022decades, claro2019multilingual}, as well as text documents coming from different domains \cite{ariyanto2024systematic} or written using various perspectives and writing styles \cite{abdullah2023systematic}. Additionally, several voices have warned against the naive use of social data, or any kind of digital trace produced by or about people \cite{olteanu2019social}, as methodological limitations such as the lack of \emph{construct validity} (i.e., whether a specific measurement, like the amount of authored documents, actually measures the theoretical construct of interest, such as human expertise) and biases introduced along the analysis pipeline are often overlooked, leading to important negative implications \cite{olteanu2019social, barocas2016big, blodgett2020language}. More recent AI developments, such as LLM-based methods, have been shown to introduce additional risks via the fabrication of inaccurate information, generating factually incorrect data by extrapolating information from the biases in training data, misinterpreting ambiguous prompts or modifying the information to align superficially with the input \cite{ji2023survey, tonmoy2024comprehensive}. In enterprise knowledge systems, emerging LLM-based approaches can exacerbate and pose new risks \cite{gausen2023framework}, introducing biases into existing information retrieval capabilities \cite{dai2024bias, bommasani2021opportunities} and magnifying existing risks such as exposure inequality \cite{fabbri2022exposure, amendola2024understanding, singh2018fairness, anthis2024impossibility}. As connections between knowledge, artefacts and people are automatically extracted and surfaced, important challenges at the intersection of work and the social also arise, such as the introduction of new workloads for those individuals linked to knowledge areas in the system or the reinforcement of existing power dynamics in an organization \cite{larsen2022ethical}. Therefore, as compared to non-AI approaches, the introduction of AI methods in sensitive contexts like knowledge management systems opens the space to new technical and societal challenges, as inaccuracies in knowledge extraction and surfacing can lead to disproportionate harms \cite{gausen2023framework}, requiring a holistic understanding of the AI's sociotechnical impacts \cite{solaiman2023evaluating}.

While past work has highlighted the increased risks to workers posed by LLMs and people recommenders recognizing transparency as a key dimension of responsible design \cite{gausen2023framework, larsen2022ethical, fabbri2022exposure, ferrara2022link, castillo2019fairness}, to the best of our knowledge, related work on how to approach transparency in AI knowledge systems that represent people is scarce. In recommender and LLM-based systems, past research has recognized the closed-box nature of deep AI methods as a key barrier towards achieving transparency \cite{zhang2020explainable, liao2023ai}. Motivating transparency from a trust and user acceptance angle, past works in social recommender systems have highlighted the importance of not only explaining why certain items are recommended by an AI algorithm, but also integrating controllability with transparency, increasing end-user satisfaction by enabling end-users to participate in the recommendation process \cite{tsai2021effects, han2013supporting, tsai2018beyond}. As such, past works have integrated various kinds of controllability inputs such as adjustments of recommendation preferences \cite{tsai2021effects}, in combination with explainability measures of familiarity, social closeness and commonalities between people as a way to explain both people \cite{guy2009you} and item recommendations \cite{bonhard2006knowing}. Recent work has also proposed making visible users’ algorithmic representations in recommendation and personalization systems by communicating to end-users how they are being perceived by the system using tag clouds \cite{xu2023does}, automatically informed profiles \cite{barbosa2021design} or natural language explanations \cite{radlinski2022natural, christakopoulou2023large}. In the realms of knowledge management, knowledge sources (that is, documents and other resources provided as ``evidence'' for a given association) and knowledge currency (whether an association is up to date) have been proposed as transparency requirements for enterprise knowledge bases \cite{wolf2020knowledge}.

\subsection{Theories of self: self-concept and its impact on human motivation and learning}

As AI knowledge systems get embedded into organizational contexts, different ways to approach AI transparency might lead to different impacts on self-perception and social processes. But to what extent can transparency modulate these effects? And to what extent can AI knowledge systems influence how individuals perceive their identity at work, or how they evaluate their contributions? Could these effects be detrimental to workers’ well-being and productivity, going against the same goals for which these systems are developed and deployed? In this section, we start by reviewing different self-concept theories from psychology and sociology that explain how individuals form and maintain their self-perception and how the formation of beliefs about oneself can shape human behavior, motivation and well-being in the workplace. Next, we review the reported impacts of AI systems on self-perception beliefs and social processes.

\subsubsection{Identity and self-concept}
\label{subsection:identity}
Identity and self are multifaceted. How someone thinks about, perceives or evaluates themselves is influenced by the imagined opinions and judgement of others \cite{cooley1902looking, james1984psychology, mead1934mind}, being constantly shaped by social interactions, perceived membership in a social group \cite{tajfel1978social, major1993social, crocker1989social} and expected audiences \cite{tice1992self}. Self-concept beliefs are also modulated by behavior \cite{gecas1983beyond}. As an individual learns how to best interact with an environment to achieve their goals, they form self-efficacy beliefs that influence their motivation and drive to further adapt and learn, shaping individual's confidence in their ability and capacity to accomplish a task \cite{bandura1977self}. How to best interact within an environment also means how to best interact with others and, in doing so, how to best present oneself. As if it was a continual performance \cite{goffman2002presentation}, individuals are driven to match their representations to the imagined audience’s expectations and to their ideal self. In organizational settings where goals are tied to economic incentives, the self becomes constrained and shaped by organizational discourses of power (``crystallized'' as described by Tracy et al. \cite{tracy2005fracturing}). These constraints might induce individuals to tone down disfavored attributes to fit into a given discourse, a phenomenon known as ``covering'' \cite{yoshino2006pressure, goffman2009stigma}. At work, self-perceptions of abilities, skills and knowledge become key components of self-concept beliefs \cite{leonard1999work}. Such beliefs are fundamental attributes in identity management \cite{christiansen1999defining} and are constantly shaped by collected feedback. Organizational behavior, hence, becomes driven by a desire to collect feedback that confirms and enhances existing self-perceptions and by the resolution of unpleasant states of cognitive dissonance when receiving feedback that disconfirms existing perceptions of self \cite{leonard1999work}. As individuals receive feedback from others, their meta-perceptions -- beliefs about how they are perceived by others -- are also calibrated \cite{grutterink2022thinking}. Extensive research indicates that individuals’ meta-perceptions can have implications on their affect, cognition, behavior and relationships \cite{grutterink2022thinking}. In organizations, meta-perceptions of expertise, in particular, have been shown to have deep implications for employee motivation and performance, shaping productivity, intent to stay in an organization and psychological well-being \cite{grutterink2022thinking, grutterink2013reciprocal, meister2023feeling}.

\subsubsection{Impacts of algorithmic feedback on self-perception and identity formation}
Feedback doesn’t have to be human to influence self-perceptions. Exposure to algorithmic bias in the form of personalized recommendations can shape self-perceptions of confidence and leadership skills \cite{french2018algorithmic} as well as people’s understanding of their identities and their orientation to others \cite{lee2022algorithmic, bhandari2022s, summers2016audience, ionescu2023tiktok}. The extent to which algorithmic feedback can affect self-perceptions has also been shown to be dependent on how individuals perceive causes in their environment and the extent to which system outputs are perceived to reflect the individual's intrinsic characteristics, such as their personality or interests, as opposed to external circumstances such as the behavior of other people in the platform or software updates \cite{french2018algorithmic}. How algorithmic inferences are interpreted is heavily dependent on how people make sense of complex systems and the informal theories they establish to explain what they see and experience while using the technology. Known as \emph{folk theories}, these theories are influenced by prior beliefs and expectations \cite{devito2017algorithms}, are malleable and dynamic \cite{devito2018people}, can be influenced by the system’s transparency \cite{eslami2016first} and evolve with direct experience, helping individuals navigate the complexity of ever-changing AI systems. 

As AI-mediated technology gets embedded into social processes, folk theories and algorithmic imaginaries enable individuals to cope with social boundaries that become blurry and distorted, helping them navigate spaces where audiences aren’t directly seen but rather imagined and collapsed into one \cite{marwick2011tweet}. For example, social media technologies such as X (formerly Twitter) flatten multiple audiences into one, a phenomenon known as ``context collapse'' that forces end-users to present themselves in the same way to different audiences or people in their network \cite{marwick2011tweet}. In online environments where the way an individual gets represented can impact both social capital and economic opportunities, user understandings of social media systems can influence their ability to adapt to them, driving self-presentation behavior \cite{devito2018people} and mediating self-transformative effects \cite{gonzales2008identity}. As AI systems represent people, individuals may have little control over representations that are derived from having done something at a particular point in time and/or from sharing attributes with a particular demographic segment, prompting those being represented to adopt different strategies in an effort to resolve dissonances between their algorithmic selves and their perceived and/or ideal selves, such as creating distinct online identities or embracing sites likely to appear high in rankings in self search results \cite{marshall2014searching}, adjusting work strategies in the case of gig \cite{lee2015working} and digital workers \cite{choi2023creator} and being deliberate in how to interact with technology in social media apps \cite{lee2022algorithmic}.

As we design and develop systems aimed at improving knowledge workers’ well-being and productivity, thinking through the impacts of technology on self-concept beliefs becomes of paramount importance. As we assess different scenarios and ramifications below, we see that unintended transparency effects might counteract core system purposes, potentially leading to worse-off situations at the individual level.

\section{Transparency in Enterprise AI Knowledge Systems}

The literature review highlighted a number of important considerations when addressing transparency in AI knowledge systems. First, transparency is multifaceted and closely linked to \emph{seeing} and perception. It determines what information is revealed or concealed, impacting what users perceive while interacting with AI systems. Second, the use of AI algorithms to represent individuals and their expertise in enterprise knowledge systems introduces ethical risks, as the opacity of these algorithms, poor performance on unseen data and lack of construct validity can lead to representational, allocative and interpersonal harms. Transparency mechanisms are necessary not only to mitigate these risks but also to ensure end-users can understand and trust the system's representations. Lastly, past research has shown that AI systems embedded in organizational and social contexts can influence individuals' self-perception and identity formation, with these effects being shaped by transparency mechanisms and users' interpretation of system outputs.

Drawing from these insights, we hypothesize that transparency in AI knowledge systems can shape different ways of seeing in the workplace. Thus, we start our analysis of transparency requirements in AI knowledge systems by assessing the types of \emph{seeing} enabled by these systems in organizational settings. To identify these, we use the looking-glass metaphor as a way to conceptualize AI knowledge systems, bringing to the foreground the social context surrounding their use. From this perspective, we identify different forms of understanding sought by individuals in organizational settings, motivating the need for different transparency dimensions: system transparency, procedural transparency and transparency of outcomes.

\subsection{The looking-glass metaphor: seeing \emph{into} and \emph{through} AI knowledge systems}

\begin{figure}[h]
  \centering
  \includegraphics[width=\linewidth]{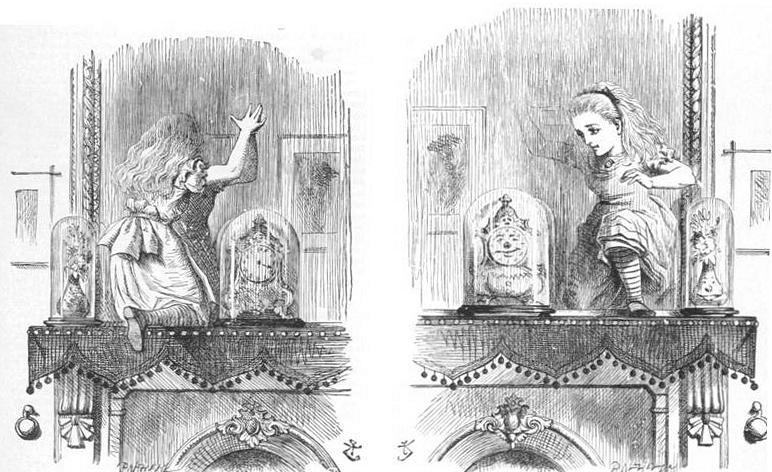}
  \caption{Alice entering the looking-glass. Illustration by John Tenniel. \\
  (Source: \href{https://en.wikipedia.org/wiki/Through\_the\_Looking-Glass\#/media/File:Aliceroom3.jpg}{https://en.wikipedia.org/wiki/Through\_the\_Looking-Glass\#/media/File:Aliceroom3.jpg})}.
  \label{fig:alice}
\end{figure}

\begin{figure}[t!]
    \subfloat[Seeing into the system]{
        \includegraphics[width=.42\linewidth]{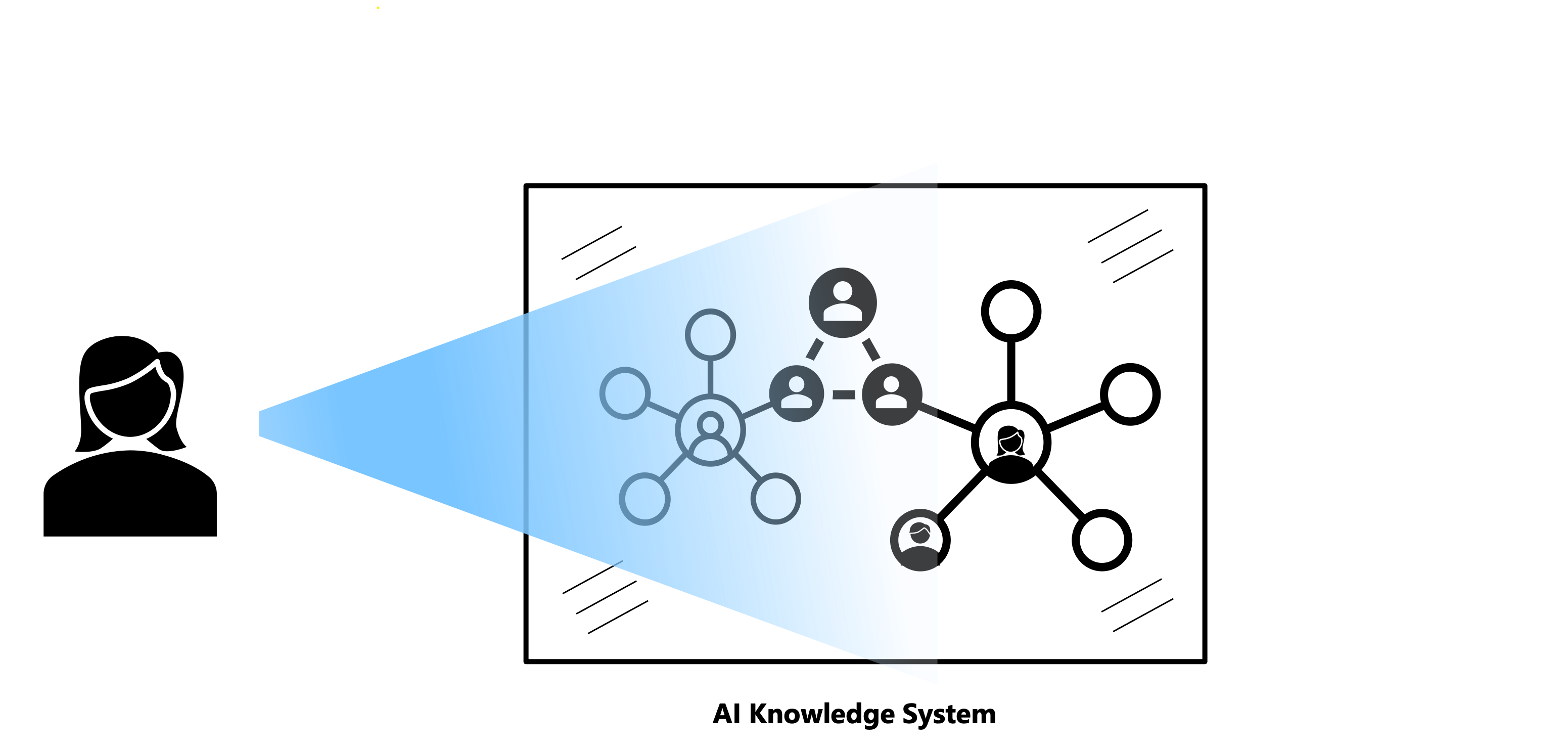}
        \label{fig:seeing-into}
    }\hfill
    \subfloat[Seeing others individually through the system]{
        \includegraphics[width=.48\linewidth]{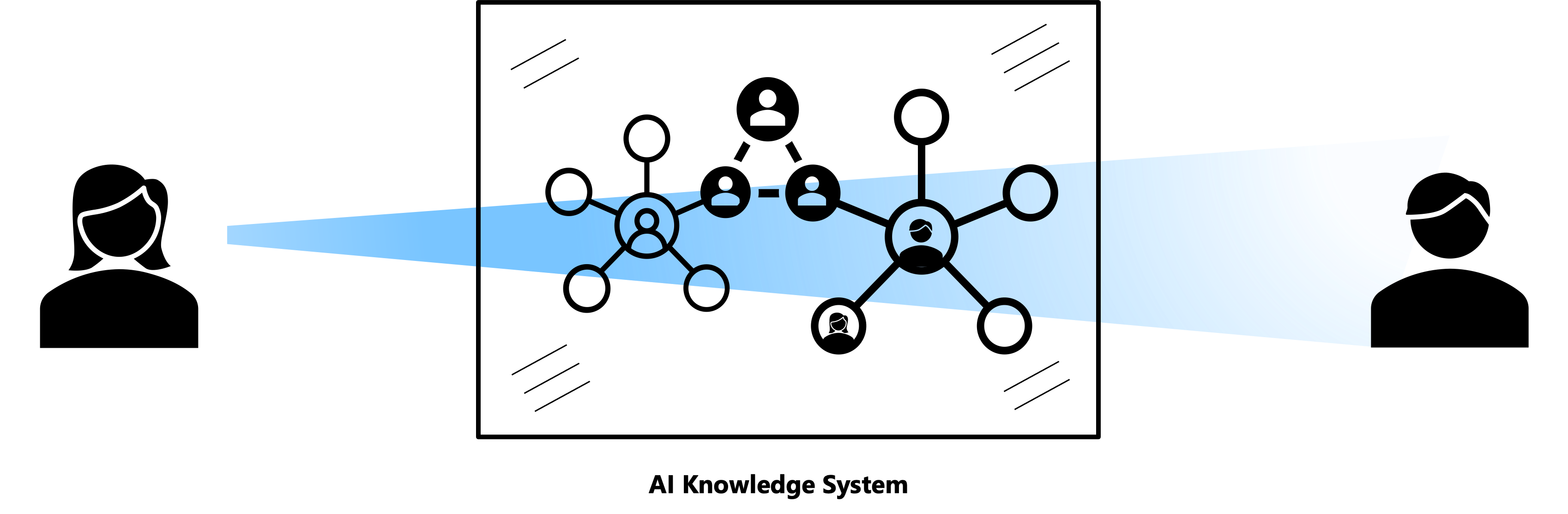}
        \label{fig:seeing-oi}
    }\\
    \subfloat[Seeing others collectively through the system]{
        \includegraphics[width=.48\linewidth]{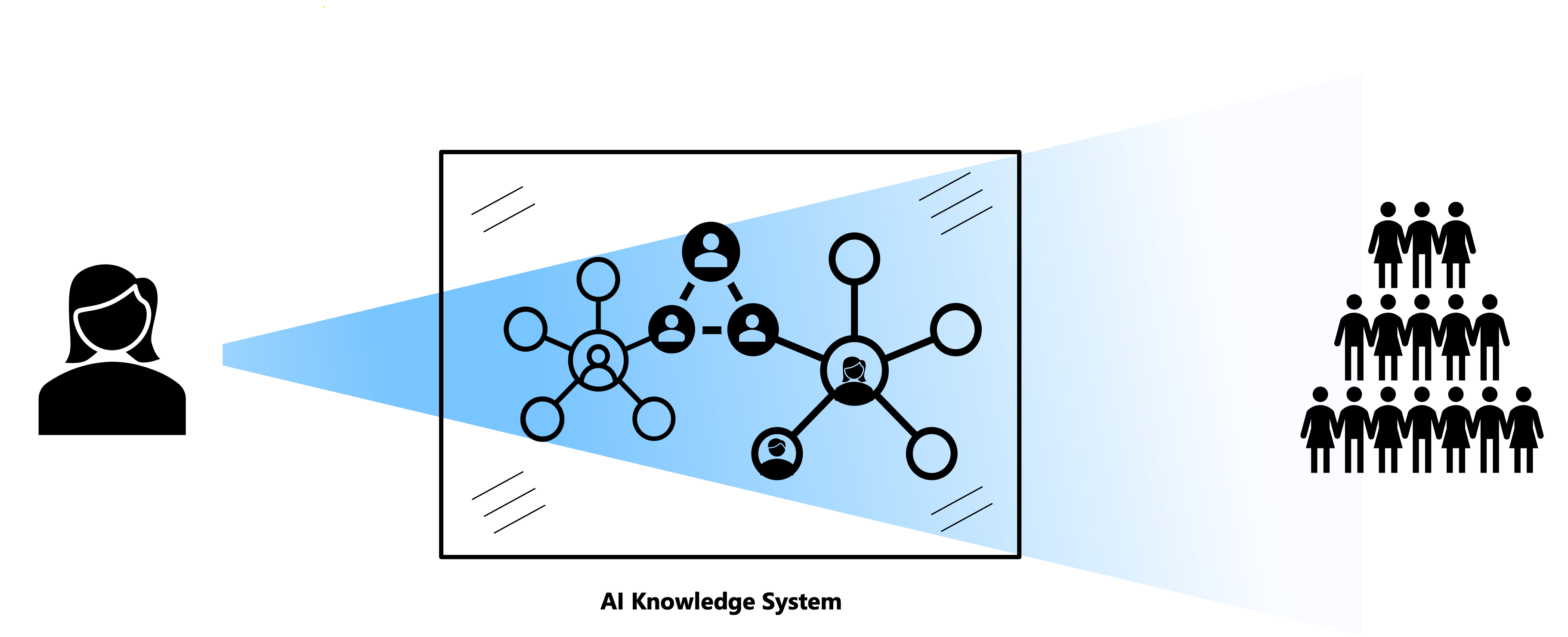}
        \label{fig:seeing-oc}
    }\hfill
    \subfloat[Seeing self through the system]{
        \includegraphics[width=.48\linewidth]{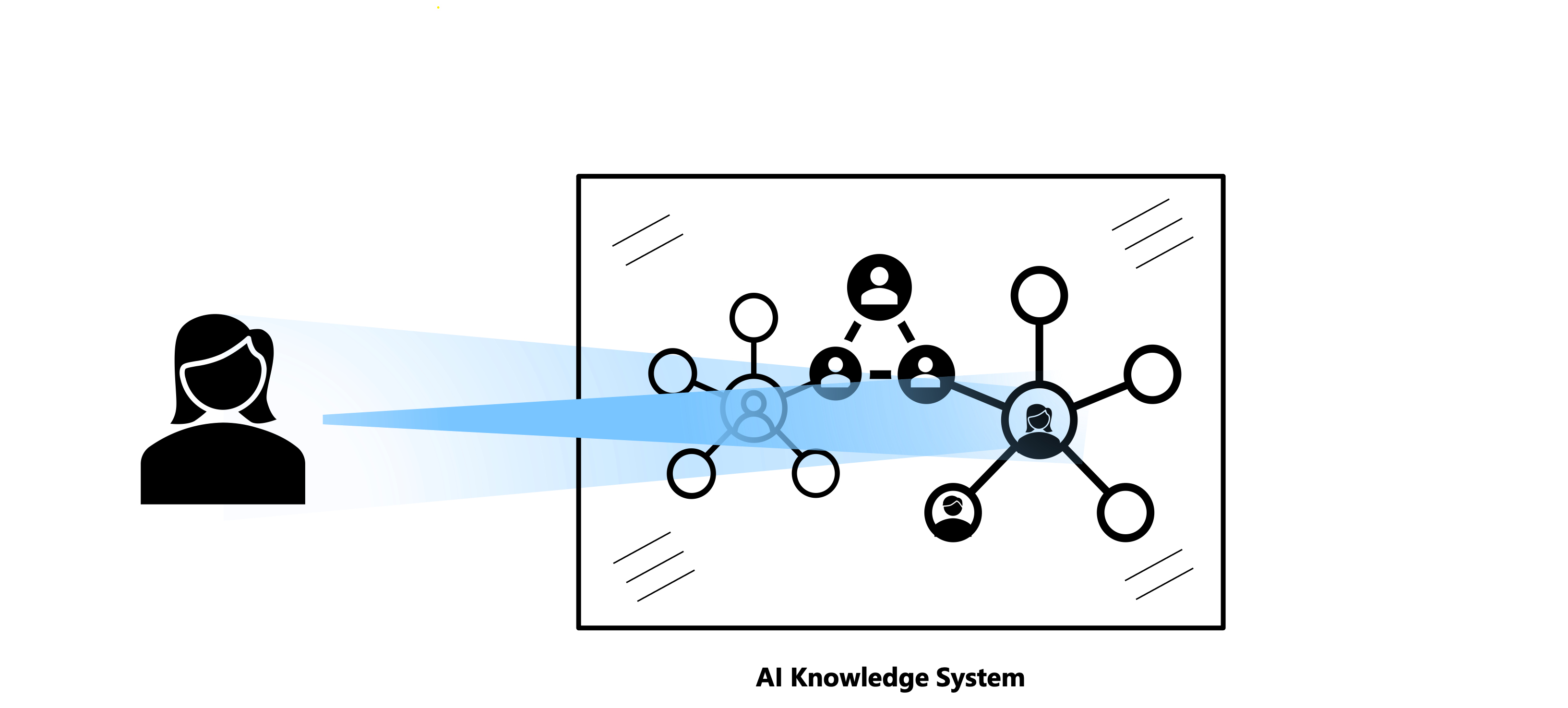}
        \label{fig:seeing-self}
    }
    \caption{When people interact with AI knowledge systems, they experience different ways of seeing. In the visual representations above, the blue lines illustrate these forms of perception. Transparency mechanisms can influence these types of seeing, shaping how users perceive the system, others and themselves.}
    \label{fig:waysofseeing}
\end{figure}

In the sequel to Alice's Adventures in Wonderland \cite{carrol2006through}, Carroll depicts a magical mirror or \emph{looking-glass} through which Alice enters a strange world that is recognizable, yet subtly altered. When Alice sees herself and steps through this mirror, she finds herself in a parallel reality where things get distorted (Figure~\ref{fig:alice}). Similarly, as AI knowledge systems get embedded into the workflows of knowledge workers, individuals might be confronted with different reflections of their own work. These reflections might not be accurate or aligned with people's own perceptions of their contributions or the way they wish to portray themselves to others in the organization. As they engage with algorithmic representations of people and knowledge, they might step into new settings where views of others are simplified and distorted (Figure~\ref{fig:waysofseeing}). Inspired by past literature depicting AI systems as mirrors \cite{french2018algorithmic, hess2014you} and crystals \cite{lee2022algorithmic}, we use the looking-glass metaphor to conceptualize enterprise AI knowledge systems as systems that both reflect and distort, allowing individuals to see themselves and others via algorithmic lenses. As compared to the mirror metaphor \cite{french2018algorithmic, hess2014you}, the looking-glass metaphor conveys the idea that AI systems not only reflect, but also distort and reshape what is perceived, often in ways that are imperceptible to users. Similarly, while the crystal metaphor implies that AI systems can shape perspectives of others as refracted images of the self (e.g., by suggesting other people with similar interests in social media platforms) \cite{lee2022algorithmic}, the looking-glass metaphor emphasizes how AI systems distort reality in ways users are unaware of, leading them to accept skewed perceptions as truth.
Using this conceptual frame, we identify different types of seeing enabled by these systems and reflect on the questions end-users may pose while \emph{seeing}.

\subsubsection{Seeing \emph{into} the system} 
If Alice knew how the looking-glass worked, this understanding would have probably shaped the way she experienced the alternative world she stepped into. AI knowledge systems that represent people and their work often consist of multiple interacting models, services and components. While some of the AI models used to extract and surface associations might be opaque, like LLMs, other components relying on traditional AI approaches, such as enterprise knowledge graphs, allow for some degree of transparency and insight into system behavior. Thus, even if full algorithmic transparency might not be possible, some of the inner workings of the system and the decisions made during its design and development can be exposed through documentation and user interface explanations, allowing users to see \emph{into} the system. For example, users might want to know what type of data is mined when extracting knowledge (such as written documents) or what metrics are used to rank people associated with a knowledge area (e.g., the number of authored documents). If we take the perspective of an end-user facing an AI knowledge system with opaque behavior, additional questions about the system might arise. For instance, how is the data used to extract knowledge? In which organizational contexts is this data being created? What aspects of work are impacted by these data extraction practices and how might these influence system drift and adaptations over time? As knowledge is surfaced to end-users, questions might also cover what are the contexts in which knowledge is surfaced and how the exposure to algorithmic representations shapes knowledge sharing practices in the organization over time.

\subsubsection{Seeing others individually \emph{through} the system} 
Because AI knowledge systems automatically extract and surface knowledge using data that is directly or indirectly produced by people as they work, individuals see both the extracted knowledge and, through that, the people who created the data. For example, in systems that mine associations between knowledge areas and knowledge artefacts \cite{vivatopics, glean}, users might be able to see who are the authors of artefacts being surfaced (Figure~\ref{fig:associations}). In systems that surface explicit associations between areas of expertise and knowledgeable individuals \cite{vivatopics, glean, bloomfire}, end-users can directly relate a given knowledge topic to an individual (Figure~\ref{fig:associations}). As people use the system and step through the mirror, the exposure to knowledge created by others \emph{and} extracted by the system can modulate how they see other individuals within their organization, allowing them to infer constructs such as expertise. As AI algorithms often under-perform in low-resource domains and can only mine explicit, externalized knowledge (Section \ref{subsection:risks}), not all work generated in an organization might be mined and surfaced. This can lead to a distortion in how individuals are represented and seen in the organization, implicitly inducing proxies for expertise and indirectly shaping interpersonal trust.

When stepping through the mirror, the pieces of work within the system's training domain might be given a better light, while unseen examples outside this domain, which are harder to extract, may be overlooked. In this way, people who create knowledge in forms suited for typical AI extraction, such as individuals authoring textual documents, might be more visible than those whose contributions are in a form less interpretable by current AI systems. As a simple example, the system may under-perform in low-resource languages due to the comparatively low number of examples it can draw on. In an English-speaking organization, a Japanese subject matter expert might not get exposure if their working documents are authored in Japanese, while another individual who is less knowledgeable about the topic but writes about it in English might be identified as an ``expert'', potentially leading to new, undesirable workloads for this individual while overlooking the actual expert. This example illustrates a sociotechnical problem, where the technical limitations of the system in low-resource languages intersect with organizational dynamics, leading to unequal distributions of work and misrepresentation of expertise within the organization.

As end-users see others individually \emph{through} the system, understanding how the system extracts the exposed associations can help them decide how to best act on the system outputs. For example, what data features and documents does the system consider when relating someone else to an area of knowledge? How reliable are the associations between a given individual and the knowledge identified by the system? What actions and knowledge contributions support these associations and how aligned are algorithmic representations of a person's expertise with the way they see their own competences and skills? Are these individuals aware of these algorithmic representations and do they feel comfortable being contacted about the knowledge areas with which they are associated? As the system pushes to the foreground knowledge artefacts relating people and knowledge, how does this shape perceptions of expertise across the organization? And how does this influence how users interact and collaborate with others?
    
\subsubsection{Seeing others collectively \emph{through} the system} 
As end-users see others individually in the organization \emph{through} the AI knowledge system, these experiences might accumulate over time, implicitly informing people's schemas of others and contributing to the establishment of patterns and categorizations that help them navigate organizational complexity and the ever-changing dynamics of work. As seen in Section \ref{subsection:risks}, issues of construct validity when quantifying human expertise can cause AI knowledge systems to inadequately recognize different types of expertise, entirely overlooking the expertise of certain groups and potentially reinforcing stereotypes. When individuals step through the algorithmic \emph{looking-glass} they might only see certain segments of people in their organization \cite{bowker2000sorting}. For example, demographic groups working with external collaborators or with lower tenure in the organization might not be seen, despite the knowledge they hold and help create. The variety of data sources considered by AI knowledge systems and the importance given to proxies that are easy to quantify, such as the amount of documents or edits, directly determines which segments get exposure to their work and which groups see their labor systematically erased.

As end-users step \emph{through} the mirror and see others collectively through the system lenses, several questions might arise. For instance, who gets to be seen and which groups are systematically erased? How do the data types considered by the system and data features extracted influence which pieces of work are seen? Are there groups in the organization that contribute to knowledge not captured by the system? What incentives are in place driving knowledge contributions suitable for system extraction? How does the system influence the exposure to different segments of people over time and what biases and proxies of expertise are being introduced?

\subsubsection{Seeing self \emph{through} the system} 
As individuals see others \emph{through} the system, they might wonder how they are seen themselves by others in their organization. In doing so, they might wish to know how they're seen by the system, which pieces of work are accessible and not accessible to the system and how their knowledge contributions are made visible to others. Drawing on the literature review, this type of seeing has the potential to impact self-concept perceptions, affecting self-efficacy beliefs when the system fails to correctly recognize knowledge contributions. As individuals see themselves through the system, the reflections might be distorted and not accurate of what others see. For instance, since the view of a person depends on what data sources the observer has access to, an individual may be represented based on documents they themselves don't have access to, leading to misalingments between how they perceive themselves and how they're seen by others through the system. As shown in Section \ref{subsection:identity}, when these reflections aren't aligned with an individual perceived and/or ideal self, individuals might take action to solve potential dissonances. As an example, if the system surfaces a list of individuals related to a knowledge area indicating the amount of documents they've authored (as illustrated in Figure~\ref{fig:associations}), when the system misses a given individual working on the topic, this individual might increase the amount of documents they author to get exposure in the system, aligning system outputs with the way they wish to be represented. As AI knowledge systems rely on quantifiable aspects of externalized knowledge, distorted reflections placing disproportionate weight on simple metrics can lead to perverse incentives, prompting changes in behavior that are not necessarily aligned with productivity nor employee well-being. In particular, missed knowledge and wrong surfaced associations can lead to representational harms, inducing feelings of alienation and lack of social control.

As end-users are confronted with these reflections, they might seek understanding about how they're seen by the system and how they're seen by others \emph{through} the algorithmic lenses. From a system perspective, questions might include what data sources and data features are considered by the system when extracting a given association, why might algorithmic associations be incorrect or incomplete and how can end-users see and control these associations. As associations are surfaced to others, who can see these and in what way? How are system outputs used by others in context? As the system allows individuals to know and control how they're seen, do people in the organization have different levels of visibility and/or control (e.g., depending on their role or hierarchical position)? As an outcome of seeing and interacting with system reflections, how is the system shaping an individual's behavior towards knowledge creation and sharing and how might these changes in behavior influence system adaptations over time?

\subsection{Transparency requirements in AI knowledge systems}
While the list of questions above isn't exhaustive, thinking about these questions can help us discern different forms of transparency needed to help end-users interpret what they see while interacting with AI knowledge systems. Some of these questions relate to \emph{technical} information about the features implemented by the system and the behavior of AI mechanisms used to extract and surface knowledge. Other questions, however, relate to \emph{social} information about the production and use of knowledge in the organization and the way the system is embedded into knowledge workflows and practices. As the \emph{technical} and the \emph{social} interact, different outcomes at the individual and organizational level might take place. Based on this, we categorize the questions and transparency requirements into three key dimensions:

\begin{itemize}
\item \textbf{System transparency:} information about the AI knowledge system targeting how the system works, including what data is consumed by the system and how knowledge gets extracted and surfaced via AI. This type of transparency helps end-users understand when they're seeing things \emph{through} algorithmic lenses and how these ways of seeing might be incomplete, distorted and governed by system rules and behavior.
\item \textbf{Procedural transparency:} this transparency dimension covers how the system operates \emph{in context} and how the system is embedded in the organization. This includes how system outputs are used and consumed by people in different organizational contexts, the norms surrounding system use as well as the processes and practices established by the organization when adopting and deploying the technology. 
\item \textbf{Transparency of outcomes:} making visible the impacts and outcomes resulting from the \emph{looking-glass} use is fundamental to help end-users understand how the system impacts themselves and their organization in the long run, how their behavior might be influenced as they use the system over time and how these changes might in turn affect system development and adaptations.
\end{itemize}

As individuals make use of AI knowledge systems, they see not only \emph{into} the system, but also others and themselves \emph{through} the system. When we think of AI knowledge systems as \emph{looking-glasses}, opening a closed-box to reveal the system's inner mechanisms ceases to be the sole focus of transparency. The looking-glass metaphor thus complements the closed-box, moving us from a narrow system focus where the primary goal is allowing end-users to \emph{see into} the system \cite{ehsan2022human} to a wider view where what lies around the system is brought to light. As we extend the discussion of transparency to span system behavior, contextual use and outcomes, different forms of understanding can be provided, potentially mitigating representational, allocative and interpersonal harms by helping end-users better interpret and make sense of what they see. Table~\ref{tab:questions} outlines questions practitioners may ask themselves to interrogate the intersections between ways of seeing and the three dimensions of transparency. However, what we should implement from an end-user and social perspective might not be aligned with what we can technically implement in practice. In the following section, we bring light to this sociotechnical gap and identify key challenges arising in the implementation of these forms of transparency.

\begin{table}[htp]
    \centering
    \caption{Questions for interrogating the intersections between ways of seeing and the three dimensions of transparency.}
    \label{tab:questions}
    \footnotesize
    \begin{tabularx}{\textwidth}{>{\raggedright\arraybackslash}p{0.15\linewidth} X X X} 
      \toprule
       & System transparency & Procedural transparency & Transparency of outcomes\\
      \midrule
      \textbf{Seeing \textit{into} the system} &
        \begin{itemize}[label={}, leftmargin=*]
        \setlength\itemsep{1em} 
            \item What type of data is mined by the system and how is it used to extract knowledge?
            \item How does the system surface knowledge? E.g., what metrics does the system use to rank knowledge contributors, and how does this influence where they are surfaced or how they are presented?
        \end{itemize} &
        \begin{itemize}[label={}, leftmargin=*]
            \setlength\itemsep{1em} 
                \item In which organizational contexts is data created? Which of the organizational data is the system using? 
                \item In which organizational contexts is knowledge being surfaced?
            \end{itemize} &
        \begin{itemize}[label={}, leftmargin=*]
        \setlength\itemsep{1em} 
            \item What aspects of work are impacted by data extraction practices? How might altered work practices and outputs influence the system over time?
            \item How is the system shaping knowledge sharing practices in the organization?
        \end{itemize} \\
      \midrule
      \textbf{Seeing others individually \textit{through} the system} &
      \begin{itemize}[label={}, leftmargin=*]
      \setlength\itemsep{1em} 
          \item What data and features does the system consider when relating someone to an area of knowledge?
          \item How reliable are surfaced associations between a given individual and knowledge?
      \end{itemize} &
      \begin{itemize}[label={}, leftmargin=*]
          \setlength\itemsep{1em} 
              \item What actions and knowledge contributions in the organization support the associations inferred by the system? Do these reflect the way expertise is established socially in the organization?
              \item How comfortable are people with the way they are being associated with knowledge areas? E.g. are they open to be contacted about the areas they have been associated with?
          \end{itemize} &
      \begin{itemize}[label={}, leftmargin=*]
      \setlength\itemsep{1em} 
          \item How is the system shaping the way the expertise of individuals is perceived in the organization?
          \item How is the system shaping the way individuals interact with others in the organization over time?
      \end{itemize} \\
      \midrule
      \textbf{Seeing others collectively \textit{through} the system} &
      \begin{itemize}[label={}, leftmargin=*]
      \setlength\itemsep{1em} 
          \item How does the type of data considered by the system influences who gets to be seen?
          \item What features and metrics does the system consider, and how do these influence the pieces of work and contributors that end-users see?
      \end{itemize} &
      \begin{itemize}[label={}, leftmargin=*]
          \setlength\itemsep{1em} 
              \item Are there groups of people in the organization who contribute knowledge that is not captured by the system?
              \item What incentives in the organization drive knowledge contributions that are suitable for system extraction?
          \end{itemize} &
      \begin{itemize}[label={}, leftmargin=*]
      \setlength\itemsep{1em} 
          \item What are the impacts of disparate system performance across different demographic subgroups? How is the system shaping the exposure to different segments of people over time?
          \item How is the system shaping the establishment of objectives and measures of success in the organization?
      \end{itemize} \\
      \midrule
      \textbf{Seeing self \textit{through} the system} &
      \begin{itemize}[label={}, leftmargin=*]
      \setlength\itemsep{1em} 
          \item What data and features does the system use to represent an individual's work? Why might algorithmic associations be incorrect and/or incomplete?
          \item How can end-users know and control how they're represented by the system?
      \end{itemize} &
      \begin{itemize}[label={}, leftmargin=*]
          \setlength\itemsep{1em} 
              \item How well does the surfaced system representation match individuals’ self-perception?
              \item How can end-users know and control how they're seen by others through the system? Do people in the organization have different levels of visibility into and/or control of how they're seen?
          \end{itemize} &
      \begin{itemize}[label={}, leftmargin=*]
      \setlength\itemsep{1em} 
          \item How are individuals adapting their behavior towards knowledge creation and sharing to bridge their self-perception and the system representation?
          \item Can individuals tell how changes in their behavior influence system behavior over time?
      \end{itemize} \\
      \bottomrule
    \end{tabularx}
\end{table}

\section{Transparency in enterprise AI knowledge systems: key challenges and implications}

How transparency gets implemented matters. In this section, we explore what challenges might arise when implementing different forms of transparency in AI knowledge systems. Thinking about these challenges is important to bring into light the gap between what we can technically implement and what we need to implement from a social perspective. When thinking about potential challenges, differentiating between system transparency, procedural transparency and transparency of outcomes can help us identify which aspects require a technical solution and which ones require a social approach, informing further interdisciplinary discourse and decision-making in work settings. As the use of emerging AI techniques in enterprise knowledge systems raises multiple challenges, introducing new risks and magnifying existing ones (Section \ref{subsection:risks}), our discussion is focused on enterprise knowledge systems powered by AI. Still, we recognize that some of the issues covered here are not unique to AI and can also occur with non-AI approaches such as heuristic rules.

\subsection{System transparency}
Several technical challenges arise in the implementation of system transparency in AI knowledge systems. While some AI methods such as deep learning techniques are inherently opaque, in this section we discuss the fundamental challenges arising with basic mechanisms of transparency, such as exposing the data used for inference and knowledge extraction. We acknowledge that in the presence of opaque algorithms, these mechanisms are not sufficient to make AI knowledge systems fully transparent and that the issues discussed in this section can still arise if transparent and non-opaque AI algorithms were used for knowledge extraction. In what follows, we explore some of the difficulties related to documentation and communication of system behavior, faithfulness, actionability and perverse incentives in work settings.

\begin{itemize}
    \item \textbf{Communication of system capabilities, mechanisms and limitations:} Especially in the age of LMMs, the way information gets communicated matters \cite{liao2023ai}. When communicating system capabilities and limitations, special care must be taken to account for different backgrounds and levels of exposure to AI. As different sources of information and media continue to shape folk theories about AI, public perception of AI technology is likely to include misconceptions, limiting the extent to which end-users can understand and act upon system documentation \cite{sartori2023minding, bewersdorff2023myths}. Due to the proprietary nature of many commercialized AI systems and competitive factors, providers have incentives to withhold detailed information about how and where AI is being used, limiting the amount of information that is shared with the public. When detailed information is provided, system documentation might be ignored, misinterpreted or used for the benefit of certain stakeholder groups over others. For instance, when communicating non-intended uses, the exception can become the rule if the consumer is inadvertently prompted to misuse the system. In this way, documenting that the system might offer only a partial view of an employee's contributions could highlight the fact that the system does provide \emph{some} insights into employee performance and productivity, directing the consumer's attention to that potential misuse. To prevent uses that shouldn't be supported by the technology, full transparency is, hence, not enough and can, in fact, be detrimental, spurring actors to behave in an adversarial manner and game the system to their advantage. Safeguards against these uses must be implemented and deployed. For instance, by preventing aggregated views and their use in reporting on people, or by incorporating inverse transparency by design \cite{zieglmeier2023rethinking}.
    \item \textbf{Faithfulness and access:} One way to improve the transparency of AI knowledge systems involves making visible the pieces of work used to relate people to knowledge (sometimes referred to as ``evidence''). This data might be created by different individuals belonging to different teams and sub-organizations, and since many organizations have file access rules and privacy settings that vary depending on membership of departments and teams, people's access to sources of information may vary from person to person. In practice, this means that evidence for knowledge associations may not be accessible to the end-users who see them. This can have different implications for how individuals see \emph{into} the system and how they see others and themselves \emph{through} the system.
    To illustrate the impacts when seeing \emph{into} the system, in global organizations, legal requirements and restrictions can prevent data transfer across geopolitical boundaries. This leads to differing configurations of access control that might prevent people from seeing knowledge contributions or documents authored by people in a different location than their own. If these knowledge contributions are used to explain system associations, this results in inconsistent explanations across end-users: While those who can access the documents are in a better position to understand system behavior, this is not the case for those who can't see the documents supporting the associations.
    There are similar implications for seeing \emph{others through} the system: The inability for individuals to be sure whether views of each other are shared might accumulate, so that the collective as a whole can't be certain of the degree to which they share a view of the organization. This could negatively impact organizational identity and ability to act collectively.

    Finally, there are also implications for seeing \emph{self through} the system: In principle, the resources used to make inferences about an individual can include resources that they do not have access to, such as human resources documents or other people's emails and chats that mention the individual. The inability to expose these pieces of data prevents the individual from getting the full picture of why they're represented in the system in the way they are. As an example, an individual may have been associated to a knowledge area based on, among other things, personal chat communication between their team members as they coordinate work. The right approach to transparency is not merely a matter of providing as much transparency as possible. In this case, informing the individual that some of the evidence for the association can’t be exposed due to privacy constraints could have a deep impact on the beliefs the individual holds about how they’re being seen by others in the organization. Is their continued employment being discussed in emails among management? Has someone made a complaint about them that has been documented in a human resources file? As seen in Section \ref{subsection:identity}, how an individual imagines others perceive them can influence self-efficacy beliefs and self-esteem, directly affecting people's behavior and interpersonal relationships within the organization.
    The lack of transparency into the way others see the individual can add to this: Not only are there sources that contribute to how others see them that they can't themselves access, but certain others may actually be able to access those sources, with the effect that those other people's views of the individual are informed by resources that the individual can't access, let alone control.

    These examples show that not only can discrepant access to organizational data result in an incomplete picture of the system, but negative consequences can follow both from incorrect assumptions of completeness and from attempts to fill in the gaps of the incompleteness.
    Furthermore, the use of work-related data as ``evidence'' enforces a layer of transparency in the social context of the organization, directly or indirectly exposing information that people might have chosen not to bring attention to. But failing to provide transparency can negatively affect the productivity and well-being of employees, as it hinders their ability to understand and assess the validity of system outputs and limits their agency to exercise control over how they are represented at work.
        
    
    \item \textbf{Actionability and perverse incentives:} When shared with end-users, different pieces of information might spur different actions and incentives, some of which might hinder rather than augment individuals, limiting their ability to create, share and make sense of knowledge. For example, simple proxies and metrics used by ranking algorithms such as the amount of edits or volume of documents authored by workers might incentivize those who don't get exposure to prioritize the quantity of work contributions, as opposed to their quality. Similarly, social signal such as views or likes might incentivize individuals to create content that's easily consumed, but which may only have short-lived value for them or the organization. Such adaptations of work, aimed at fitting into the system's extractive mechanisms, might lead to irrelevant associations being mined, in turn hindering the value of the system in the long run. As we think about what information to make visible, we should ask ourselves what behavior we wish to encourage when surfacing explanations, metrics or data sources in the UI. When designing the ways users can see themselves through the system, do we want to motivate them to generate and share further contributions for knowledge extraction, contributing to system value and adoption? Do we want to help them reflect on their impact and past work? What are the behaviors we're nudging, and why are nudging those?

    Other information, such as membership of demographic groups, could also be surfaced to help end-users understand system behavior. However, extreme care is needed when making these aspects transparent, as these may introduce new harms. As shown in Section \ref{subsection:identity}, within an organization, self-identity might be shaped by the perceived membership to multiple groups at once, such as job role, seniority, as well as demographic segments defined by age, gender or location. Some of these groups might be historically marginalized groups and perceived as stigmatized. A vast volume of past psychological research has shown that beliefs of membership within different stigmatized groups and cognitions held by individuals about their stigma, including their awareness of stereotypes associated with their group, might result in unique combinations of discrimination and privilege (often referred to as intersectionality \cite{crenshaw2017intersectionality, erete2021can}), and have important affective, motivational and interpersonal consequences \cite{major1993social, crocker1989social}. Making transparent how membership within a group influences system behavior could enforce stigma beliefs, leading to perverse incentives where employees alter their behavior to fit into group expectations, potentially affecting the well-being, motivation and sense of belonging in the workplace.
\end{itemize}

\subsection{Procedural transparency}
When making transparent how AI knowledge systems are embedded in a given organization, several dimensions related to system use in context come to play. These dimensions might fall outside the system boundaries, belonging to the \emph{social}, rather than the \emph{technical}, and might not be able to be tracked and summarized in a way that's actionable and valuable to end-users:

\begin{itemize}
    \item \textbf{Tracking system use in context:} As the system automatically mines data and creates representations of individuals and their work, knowledge workers might care not only about how the system relates them to knowledge, but also about how system outputs are shown to others and used in practice. To help users achieve their goals, transparency needs to go beyond explaining the provenance of system outputs. It needs to make transparent what happens next: in which contexts are these inferences surfaced, to whom and how are these used in practice? Being transparent about how representations are used can give agency to individuals, allowing them to understand and control how they're being seen and how their work is presented to others. While tracking where representations are surfaced and to whom might be technically possible, the use of people's representations might not be explicitly documented and, hence, not within the scope of what can be made transparent. For example, whether a representation is actually being used to progress work, to establish a collaboration or to assess the work of others might not be tracked anywhere. From a symbolic interactionism perspective \cite{blumer1986symbolic}, the meaning ascribed to the associations being surfaced by AI knowledge systems, and therefore their use, might be a product of the social interactions happening outside of the scope of the system. Hence, being transparent on how representations are consumed and the norms surrounding system use in context requires the establishment of social processes in the organization, such as sharing clear guidelines for the ethical use of AI-generated representations in decision-making, developing organizational policies that prevent the misuse of these representations in certain contexts or implementing feedback channels for employees to raise concerns.
    \item \textbf{Curse of dimensionality:} As light passing through glass in infinite ways and creating different views \emph{through}, people representations in AI knowledge systems might be surfaced in numerous ways, offering different views depending on who is seeing and the context where seeing takes place. Being transparent about how people representations are presented to others and used by others opens a big space of possibilities for which information to reveal. As the number of users and contexts of use grows, the amount of information that can be made visible grows exponentially. Similarly to the exponential increase of computational complexity required in machine learning when data grows in dimensions (the curse of dimensionality), the multi-perspective nature of AI knowledge systems and the contextual variance in how extracted knowledge is (re-)presented and interpreted poses several challenges to summarizing the information in a faithful and accurate way. When summarizing and condensing information related to system use, we also need to preserve the privacy of those who see and those who are being seen. As we think about helping end-users understand how the system operates in context, communicating all the ways in which individuals are related to knowledge in the system and all the ways in which this information is surfaced to others in the organization might not be feasible nor helpful in practice, potentially leading to overwhelming amounts of information in detriment of focus and productivity.    
    \item \textbf{Organizational power asymmetries:} \citet{gausen2023framework} highlight that system opacity can make it harder for workers and data subjects to meaningfully critique the system, further intensifying existing organizational power disparities. However, transparency alone is not a one-size-fits-all solution. In the backdrop of existing power asymmetries, transparency mechanisms themselves can be misused. For instance, making visible performance metrics can contribute to workplace surveillance~\citep{teachout2022boss}, reinforcing power disparities and eroding workers' autonomy. Therefore, the design of any proposed transparency mechanism should carefully consider who is the subject and who is the consumer of the information that is being exposed, ensuring that transparency mechanisms do not disproportionately benefit those in positions of power, nor reduce workers to mere data points for organizational consumption~\citep{Ajunwa2020}.
\end{itemize}

\subsection{Transparency of Outcomes}
As we think about how AI knowledge systems could support transparency of outcomes, how to measure the impacts of using these systems in the long run involves several challenges: 

\begin{itemize}
    \item \textbf{Measuring and attributing actual outcomes:} While providers might be able to communicate intended outcomes and system goals, realized outcomes can be difficult to measure and attribute, being deeply entangled in social phenomena. For instance, changes in behavior at the individual level might respond to different incentives established within the organization as well as feedback received other than the information provided by the system. Similarly, how the system evolves over time might respond to new market needs and technological innovations, making it difficult to track those updates being influenced by changes in end-users' behavior and work-related data. Outcomes at the group level such as deviations in system performance across different demographic segments of users might be easier to track but difficult to communicate to individuals. Similarly to the negative downstream effects identified in system transparency, communicating that the system works worse for individuals belonging to a minority group could reinforce existing stigma beliefs in an organization, negatively affecting the well-being of knowledge workers. 
\end{itemize}

\section{Discussion and future directions}

Transparency is multifaceted. As we think about which forms of transparency should AI knowledge systems support, several challenges and tradeoffs emerge. In what follows, we discuss future directions and areas of transparency research in AI knowledge systems, encouraging CSCW researchers to further study transparency implications and ways to overcome implementation challenges.

\subsection{The looking-glass metaphor}
The looking-glass metaphor has been instrumental to our conversation about the reciprocal relationship between the \emph{social} and the \emph{technical} in AI knowledge systems. Whereas the closed-box metaphor directs the focus towards the system and the technical aspects that can be exposed to end-users, the looking-glass metaphor brings to the foreground the people and the social processes surrounding system use.

Considering how people become part of a system that both distorts and reflects back, and how this affects both individuals and collectives, is a vital part of understanding the personal and social impacts of AI systems. By aiding these reflections, the looking-glass metaphor has enabled our discussion about transparency to be about not just what the system does and how it does this, but also about how the system represents people and, in turn, how it causes people to see each other and themselves \emph{through} algorithmic lenses. Similar to Plato's allegory of the cave where prisoners mistake shadows for reality \cite{cornford1976republic}, the looking-glass metaphor calls attention to the risk of creating distorted perceptions of reality, where what we believe to be accurate may in fact be a skewed version of the truth. In the case of AI knowledge systems, the looking-glass metaphor can support wider discussions on how to design AI knowledge systems that augment reality in meaningful ways, rather than create distorted virtual realities that disturb knowledge relationality \cite{gausen2023framework}.

We encourage others to apply the looking-glass metaphor in other sociotechnical contexts such as social media or personalization systems and to develop reflexive questions such as the ones listed in Table~\ref{tab:questions}, to inform further exploration of transparency requirements across different dimensions while bringing attention to both the \emph{social} and the \emph{technical} aspects surrounding system use.

\subsection{System transparency}
How system transparency gets implemented matters. How information is chosen to be shared might shape in different ways the types of seeing enabled by AI systems, potentially leading to unforeseen harms. As we think about what information to make visible to help end-users understand how AI systems work, we warn against making information visible for the sake of improving system transparency or claiming the system is made “transparent”. We particularly caution practitioners and researchers against making visible algorithmic representations of people that may be incomplete and out of context in a way that might presume ground truth, such as representations curated by personalization systems \cite{radlinski2022natural, christakopoulou2023large}, without thoroughly analyzing the implications of doing so. We recommend paying special attention to the language used when communicating algorithmic inferences, using specific labels to avoid misinterpretations and avoiding making value judgements or categorizations about end-users that might be flawed. Recognizing that these systems can shape how people see themselves and how they see others, we prompt the AI community to think through and further research the effects of transparency in sociotechnical contexts, carefully assessing the ramifications of algorithmic feedback on both self-concept beliefs and the ways people perceive individuals and collectives through algorithmic lenses.

\subsection{Procedural transparency}
In theory, AI knowledge systems can also serve the needs of leaders in organizations who need to make decisions on how to best allocate their resources for enhanced productivity and competitivity in their market. Knowing areas of expertise or work within different teams and identifying overlaps within organizations becomes critical for decision making on how to best utilize and capitalize the knowledge within an enterprise context, helping global enterprises take advantage of economies of scale while minimizing the likelihood of diseconomies of scale in growing and expanding organizations \cite{andreeva2012does, coase1995nature, arrow1974limits}. In this context, AI knowledge systems might be misused in unforeseen ways, leading to interpersonal harms and loss of agency for workers, while instilling feelings of surveillance, damaging worker's well-being, productivity and enjoyment of work. In misusing the system, power dynamics in an organization might lead to withholding of information on the actual uses and norms surrounding the adoption of knowledge management technology. From a provider perspective, transparency via communication of intended and non intended uses isn't enough to prevent misuse of technology. Rather, specific guardrails need to be in place to prevent product misuse and guarantee that both workers' and organizational goals are prioritized, such as removing reporting of knowledge contributions and people analytics capabilities. System designers can also be challenged to provide power-aware transparency mechanisms that allow ``studying up''~\citep{miceli2022studying} to reduce inequities of organizational power over time. A combination of guardrails with explicit communication is also needed to prevent individual harms at scale, informing end-users of the system about intended uses of the technology as well as any potential capabilities, if any, given to other members of the organization (such as administrators).

\subsection{Transparency of Outcomes}
Even when actual outcomes might be difficult to measure, when individuals are confronted with algorithmic outcomes that are unfair and harmful, informed recourse should be provided so that those who feel adversely affected by the system know how to remediate potential unfair consequences. For example, when individuals believe system outputs are incorrect, impacting their social capital at work or generating increased labor to be reversed, clear documentation of ways to remediate potential negative outcomes should be made available. These recourse processes should be actionable and allow individuals to act on the information provided, giving them the agency to understand and control how to best present themselves through the system.

\section{Limitations}
The analysis of transparency requirements and impacts presented in this paper is reflective, value-laden and necessarily incomplete. Its aim is to encourage practitioners and researchers to critically reflect on the implications of AI transparency in sensitive contexts, rather than offering a complete account of transparency needs and potential challenges in AI knowledge systems. As foreseeing impacts that haven't been observed before is inherently a difficult task \cite{boyarskaya2020overcoming}, we introduced the looking-glass metaphor to support counterfactual thinking and encourage thoughtful transparency designs in sociotechnical contexts. While the metaphor might be applicable to other AI systems that implicitly and/or explicitly represent people, researchers might need to adjust the narrative or use the metaphor in combination with other metaphors as required by their use case. 

The analysis of transparency requirements is intended to inform further empirical research, serving as a base ground of hypotheses that require further evaluation. When access to users is limited, practitioners can use the questions outlined in \autoref{tab:questions} to start exploring the design space of transparency, focusing on how different forms of transparency can impact various ways of seeing. Similarly, the analysis of challenges presented in this paper is not exhaustive. While the analysis can help guide decision-making and inform the design of transparency mechanisms in AI knowledge systems, the challenges identified may not extend to other AI systems. Furthermore, the assessment of transparency impacts in AI systems must be iterative and ongoing, evolving alongside system development and regularly updated as new insights emerge from user interactions and feedback.

\section{Conclusion}

In this paper, we've taken a first step into identifying transparency needs, implications and challenges in enterprise AI knowledge systems. We've proposed the looking-glass metaphor as a way to conceptualize AI knowledge systems, expanding our view on transparency requirements by considering the implications of exposing information when seeing \emph{through} the system. Grounded in past research, in this paper we've formulated potential effects on self-concept beliefs and a set of reflexive questions that require further empirical and user research in work settings. As we take a step back and zoom out, understanding AI systems as systems that reflect and distort puts into light their role in the economics of information and in shaping information asymmetries. With the rise of foundation models and continuous development of AI technology, we encourage cross-disciplinary research in AI transparency to help us step back and see the bigger picture to understand its societal implications. In doing so, economic views explaining the motivations of companies creating and consuming AI knowledge systems \cite{stiglitz2000contributions}, a perspective less often sought in CSCW, might provide powerful ways to frame and put into perspective different transparency challenges and tensions arising in sociotechnical contexts. As AI systems shape new futures of work, we believe AI transparency will be critical to ensure these systems augment human labor in a way that is meaningful and equitable and we hope our discussion sparks further research.

\bibliographystyle{ACM-Reference-Format}
\bibliography{sample-base}


\end{document}